\documentstyle[12pt]{article}
\if@twoside
\else
\fi
\begin{document}
\title{ Fractional Superstrings with Space-Time\\
Critical Dimensions Four and Six}
\author{Philip C. Argyres and S.-H. Henry Tye\\
{\normalsize \it Newman Laboratory of Nuclear Studies}\\
{\normalsize \it Cornell University}\\
{\normalsize \it Ithaca, N.Y.  14853-5001}}
\date{CLNS 91/1068 \\
August 1991}
\maketitle

\begin{abstract}
We propose possible new string theories based on local world-sheet
symmetries corresponding to extensions of the Virasoro
algebra by fractional spin currents.  They have critical
central charges $c=6(K+8)/(K+2)$ and Minkowski space-time
dimensions $D=2+16/K$ for $K\geq2$ an integer.  We present
evidence for their existence by constructing modular invariant
partition functions and the massless particle spectra.
The dimension $4$ and $6$ strings have space-time supersymmetry.
\end{abstract}


String theories\cite{GSW} are primarily characterized by the local
symmetries of a two-dimensional field theory on the string
world-sheet.  The local symmetries of the bosonic string are
reparameterization and Weyl invariance which lead to a critical
string propagating in 26 flat space-time dimensions.  The
superstring enlarges the world-sheet gauge invariance to include
a local $N=1$ supersymmetry, and reduces the critical space-time
dimension to 10.  It is natural to try to construct string
theories with smaller critical space-time dimensions by changing
the worldsheet symmetry.  Unfortunately, enlarging to a
local $N=2$ supersymmetry produces a critical string in just
two-dimensional Minkowski space.  It is well known, however, that
fractional-spin fields can exist in two-dimensional field theory.
One can imagine new
local symmetries on the world-sheet which involve fractional-spin
currents and which lead to string propagation in space-times with
dimensions less than 10.  In this letter we will present evidence
for the existence of such new string theories, and in particular
will show that strings with spin $4/3$ and $6/5$ currents on the
world-sheet can have interesting phenomenologies in $6$ and $4$
space-time dimensions, respectively.

The algebra of constraints resulting from gauge-fixing the
two-dimensional local symmetry must contain the Virasoro algebra.
In this letter we consider strings with the Virasoro algebra
extended by chiral currents with fractional spins (equivalently,
conformal dimensions)
given by $\Delta=1+2/(K+2)$ for $K\geq2$ an integer.
These fractional spin currents transform bosonic fields to
spin $2/(K+2)$ fields.  We refer to these algebras
as fractional superconformal algebras and
the string theories based on these algebras as fractional
superstrings.  The $K=2$ case is the usual superstring with
the superVirasoro algebra as its constraint algebra.
The construction of fractional superstrings
generalizes that of the superstring.

In fact, the fractional superconformal algebras were
proposed a few years ago;\cite{KMQ,BNY} we can construct
them as follows.  Consider the chiral $SU(2)_K$
Wess-Zumino-Witten (WZW) theory.  Denote the WZW primary
fields of spin $j$ and $J_3$ quantum number $m$ by
$\Phi^j_m$, for $0\leq j\leq K/2$, and the $n$-th mode
of the Kac-Moody currents by $J^m_n$.  Define the
current\cite{KMQ} $\widehat{G}^K(z)=\sum_m J_{-1}^m \Phi^1_m(z)$.
The WZW theory can be expressed as the tensor product of a free
scalar field, $\phi$, compactified on a circle of radius
$1/\sqrt{K}$ and the $Z_K$-parafermion theory.\cite{ZFpara}
The parafermion fields
can be organized into sets $f^j_m$ according to $SU(2)_K$
quantum numbers.  In terms of these fields
$\widehat{G}_K$ can be expressed as
  \begin{equation}
  \widehat{G}^K(z)=\epsilon(z)\partial\phi(z)+\eta(z),
  \end{equation}
where $\epsilon$, the lowest dimension member of $f^1_0$,
is the first energy operator of the $Z_K$-parafermion
theory, and has dimension $2/(K+2)$.  The $\eta$ field (present
for $K\geq3$) is a parafermion descendant of $\epsilon$, but
Virasoro primary with dimension $\Delta$.

The chiral algebra generated by the energy-momentum tensor and
$\widehat{G}_K$ (appropriately modified with background charge
for $\phi$) was first proposed as the underlying symmetry of the
$SU(2)_K\otimes SU(2)_L/SU(2)_{K+L}$ coset models.\cite{KMQ,BNY}
Much evidence has accumulated showing the
importance of these algebras in organizing the operator
content of conformal field theories.\cite{FZft,Pog,CLT}
We should emphasize, however, that except in some
special cases\cite{FZft,AGT} the exact form of this
fractional supersymmetry algebra is not known.

Next, we replace the scalar field $\phi$ with a decompactified
one called $X$.  The $SU(2)$ WZW symmetry is now lost.
We interpret the $X$-field as the string coordinate and tensor
together $D$ copies to allow a $D$-dimensional space-time
interpretation.  Adding space-time indices, we write the
(world-sheet) fractional supersymmetry current for a
$D$-dimensional fractional superstring as
  \begin{equation}
  G^K(z)=\epsilon^\mu\partial X_\mu(z)
  +:\!\epsilon^\mu \epsilon_\mu(z)\! :,
  \label{Gkay}
  \end{equation}
where the index $\mu=0,1,2,\ldots,D-1$ is contracted with a
Minkowski metric.  The normal ordering symbol in the second term
is meant as an instruction to pick out the $\eta$ field
which appears as the dimension $\Delta$
operator in the $\epsilon\epsilon$ operator product expansion
(see also Eq.~(\ref{Gexp}) below).  Thus, the conformal field
theory corresponding to a single space-time dimension is a
$c=1$ free boson plus a $c=2(K-1)/(K+2)$ $Z_K$-parafermion theory
and has a conformal anomaly $c_0=3K/(K+2)$.
It is easy to show that the $D$-fold tensor
product algebra generated by $G^K$ closes on itself if the
single component algebra $\widehat{G}_K$ does.\cite{ALT}

For $K=2$, the $Z_K$-parafermion theory is just the
Ising model and the energy operator $\epsilon$ is simply the
free Majorana-Weyl field; the second term in Eq.~(\ref{Gkay})
vanishes in this case.  This describes the usual superstring.

We now wish to show that in addition to the $K=2$ superstring,
theories with other $K$ values can have sensible interpretations
as string theories in flat space-time.  To this end, we follow a
simple argument to determine the space-time critical dimension of
the fractional superstring as a necessary condition
for a desirable string phenomenology.  Along the way
we show that the fractional supercurrent imposes the
correct physical state condition on the massless
particles in the spectrum.

The physical state conditions arising from the
gauge fixing constraints on the worldsheet are
$G^K_n|{\rm phys}\rangle=L_n|{\rm phys}\rangle=0$
for $n>0$ and $L_0|{\rm phys}\rangle=
v|{\rm phys}\rangle$, where the $L_n$ are the
generators of the Virasoro algebra and $v$ is
the intercept.  To determine $v$ we demand that the
open string theory have a massless vector particle
(a graviton in the closed string theory).
For example, in the bosonic string, the lowest-mass
vector particle is $|\psi\rangle=
\alpha^\mu_{-1}|p\rangle$ where $\alpha^\mu_n$
are the modes of the $X^\mu$ coordinate fields
and $|p\rangle$ is the ground state with momentum $p^\nu$.
To be a physical state, it must statisfy
$0=(L_0-v)|\psi\rangle=(1-M^2-v)|\psi\rangle$.
Therefore $|\psi\rangle$ is massless only for
$v=1$.  Now apply this argument to the fractional
superstring.  For $K\geq2$ the lowest-mass vector
particle is
  \begin{equation}
  |\psi\rangle=
  \zeta_\mu\epsilon^\mu_{-2/(K+2)}|p\rangle,
  \label{Vect}
  \end{equation}
where the moding of the energy operator $\epsilon^\mu$
acting on the ground state follows from its dimension
and $\zeta_\mu$ is the polarization tensor.  The physical
state condition $0=(L_0-v)|\psi\rangle=(2/(K+2)-M^2-v)|\psi\rangle$
implies that $|\psi\rangle$ is massless when $v=2/(K+2)$.

The only nontrivial physical state condition on $|\psi\rangle$
is $0=G^K_{2/(K+2)}|\psi\rangle$.  This can be computed
using the mode expansion of the fractional supersymmetry
current following from Eq.~(\ref{Gkay})
  \begin{equation}
  G^K_{2/(K+2)}=\sum_n
  \epsilon^\mu_{2/(K+2)+n}
  \left(\alpha_{-n,\mu}+c_n\epsilon_{-n,\mu}\right),
  \label{Gexp}
  \end{equation}
where the $\alpha_n$ are the $X$-boson modes and
the $c_n$ are coefficients that can be calculated from the
$Z_K$-parafermion theory.  Since the second term annihilates
$|\psi\rangle$, we find for the physical state condition
$0=G^K_{2/(K+2)}|\psi\rangle=p\cdot\zeta|p\rangle$.
Because $|\psi\rangle$ is massless, the longitudinal state with
$\zeta^\mu\propto p^\mu$ is null.  Thus we find the physical
state condition and the $D-2$ propagating degrees of freedom
appropriate for a massless vector particle.

To be consistent with the massless vector particle,
we require that, for the whole spectrum,
effectively only $D-2$ transverse dimensions
worth of polarization states actually propagate.
This is a signal of the enlarged gauge invariance
of critical strings.

The character $\chi(q)$ which includes the string
ground state counts
the the number of propagating degrees of freedom
$N$ at each mass level $M^2$ by the coefficients
of terms in a power series expansion in $q$:
  \begin{equation}
  \chi(q)=\sum Nq^{M^2}=q^{-v}(1+\ldots).
  \label{chiv}
  \end{equation}
The second equality follows from the fact that
$L_0=-M^2+\ldots$ counts the mass level of a
state, where the dots stand for Fock space
number operators. Thus the lowest-mass physical
state obeys $M^2=-v$.  On the other hand,
conformal invariance requires that\cite{BrN}
  \begin{equation}
  \chi(q)= q^{-c_{\rm eff}/24}(1+\ldots)
  \label{chic}
  \end{equation}
where $c_{\rm eff}$ is the effective conformal
anomaly of the propagating degrees of freedom.
{}From the requirement that only the transverse
dimensions couple in a critical string, we have
$c_{\rm eff}=(D-2)c_0$.  Comparison of Eqs.~(\ref{chiv})
and (\ref{chic}) shows that $D=2+24v/c_0$.

Since $c_0=3K/(K+2)$ for the fractional superstring,
and we found $v=2/(K+2)$ above, we have for $K\geq2$
  \begin{equation}
  D=2+{16\over K},
  \end{equation}
and
  \begin{equation}
  c_{\rm crit}=Dc_0={6(K+8)\over{K+2}}.
  \end{equation}
For $K=2$, $4$, $8$, $16$ and $\infty$,
we find the new integer critical dimensions $D=10$, $6$, $4$,
$3$ and $2$, respectively.\cite{ArM}  Curiously, these
are precisely the dimensons which allow minimal super Yang-Mills
theories.  Note that for $K=2$ we recover the superstring
result $D=10$.

We will now construct modular-invariant partition
functions for the $K=4$ and $8$ closed fractional
superstrings.  As above, we impose the
conditions that only $D-2$ dimensions worth of states
propagate, and that there is a graviton in the closed
string spectrum.  In addition we
require that no tachyonic states appear in the spectrum.

Each of the $D-2$ transverse dimensions contribute a
$Z_K$-parafermion plus $X^\mu$-boson worth of states to
the partition function.  Specifically, each boson
contributes a factor of $\eta(q)^{-1}$, the reciprocal
of the Dedekind eta function, and each set
of parafermion fields $f^j_m$ contributes a factor
$\eta(q)c^{2j}_{2m}(q)$ to the partition function, where
the $c^{2j}_{2m}$ are string functions.\cite{GQ}  The
string functions obey the identities $c^{K-2j}_{K-2m}=
c^{2j}_{2m}=c^{2j}_{-2m}$, and have known power series expansions
in $q$ starting with $c^{2j}_{2m}(q)=q^{h(j,m)}(1+\ldots)$
where, for $|m|\leq j$,
  \begin{equation}
  h(j,m)={8j(j+1)-K\over 8(K+2)}-{m^2\over K}.
  \label{sjm}
  \end{equation}
The partition function must also be
invariant under the modular transformations $T:\tau
\rightarrow\tau+1$ and $S: \tau\rightarrow-1/\tau$,
where $q={\rm exp}(2i\pi\tau)$.  The $T$ modular transformation
properties of the string functions follow from
Eq.~(\ref{sjm}), while the $S$ transformation
is given by\cite{GQ}
  \begin{eqnarray}
  c^{2j}_{2m}(-1/\tau)= && \left(-i\tau K(K+2)\right)^{-1/2}
   \sum^K_{J=0} \sum^K_{M=1-K}{\rm e}^{2i\pi mM/K}\nonumber\\
  && \times{\rm sin}\left[{{\pi(2j+1)(J+1)}
   \over{K+2}}\right]c^J_M(\tau).\nonumber
  \end{eqnarray}

Let us briefly review the construction of the
partition function of the $K=2$ (superstring) theory
with critical dimension $D=10$.  The
partition function will be a sum of terms each
with eight $Z_2$-parafermion (free fermion) string
function factors for each of the eight transverse dimensions.
A tachyon-free modular invariant partition function
is found\cite{GSO} to be ${\cal Z}_{(2,2)}(q)=|A_2|^2$
where\cite{taut}
  \begin{eqnarray}
  A_2= && 8(c^0_0)^7c^2_0-8(c^1_1)^8
   +56(c^0_0)^5(c^2_0)^3\nonumber\\
  && \ +56(c^0_0)^3(c^2_0)^5 +8c^0_0(c^2_0)^7.
  \label{zeetwo}
  \end{eqnarray}
{}From Eq.~(\ref{sjm}) we see that only the first two terms
contribute to the massless spectrum, and the requirement
that the left-moving spectrum contain a massless vector
particle (or, equivalently, that ${\cal Z}_{(2,2)}$ contain
the graviton) fixes the normalization of $A_2$.
The translation to the Jacobi $\vartheta$-function
notation is given by $(c^0_0+c^2_0)^2=\vartheta_3$ and
$(c^0_0-c^2_0)^2=\vartheta_4$ for the Neveu-Schwarz sector and
$2(c^1_1)^2=\vartheta_2$ for the Ramond sector.
The massless sector in ${\cal Z}_{(2,2)}$ contains an $N=2$
supergravity multiplet and $A_2$ vanishes, consistent with
space-time supersymmetry.  We will find that all of these
features are present in the $K=4$ and $K=8$ partition
functions.

The $K=4$ fractional superstring has critical
dimension $D=6$, conformal anomaly $c_0=2$ per
space-time dimension, and a fractional supercurrent
$G^K$ of dimension $4/3$.  We find one tachyon-free
modular invariant partition function
  \begin{equation}
  {\cal Z}_{(4,4)}(q)=|A_4|^2+12|B_4|^2,
  \end{equation}
with
  \begin{eqnarray}
  A_4 = && 4(c^0_0+c^4_0)^3(c^2_0)-4(c^2_2)^4
   +32(c^0_2)^3(c^2_2)-4(c^2_0)^4,\nonumber\\
  B_4 = && -4(c^2_2)^2(c^2_0)^2
   +8(c^0_0+c^4_0)(c^2_0)(c^0_2)^2\nonumber\\
  && \ +4(c^0_0+c^4_0)^2(c^2_2)(c^0_2).\nonumber
  \end{eqnarray}
Under the $S$ modular transformation the partition
function blocks obey $S(A_4)=(1/2)A_4+3B_4$, and
$S(B_4)=(1/4)A_4-(1/2)B_4$.
Eq.~(\ref{sjm}) implies that $A_4\sim q^0(1+\ldots)$ and
$B_4\sim q^{1/2}(1+\ldots)$.  Thus we see
that there are no tachyons in this theory, and that
the only contributions to the massless states are
from the terms $4(c^0_0)^3c^2_0-4(c^4_4)^4$ in $A_4$.  The first
term has the interpretation as the massless vector
particle since it is created from the $(c^0_0)^6$ vacuum
by a parafermion field with $j=1$ and $m=0$, giving rise
to one $c^2_0$ factor.  These are precisely the quantum
numbers of the $\epsilon^\mu$ energy operators, so we
can identify it with the massless vector
state of Eq.~(\ref{Vect}).  The number
of degrees of freedom of a massless vector particle
in six dimensions is four, fixing the normalization of
the partition function.  The second term, appearing with
a minus sign, must be interpreted as a space-time fermion.
It is composed of $j=1$ spin fields in the parafermion
theory, commonly denoted $\sigma_2\in f^1_1$.  The normalization
of this term suggests that it is a space-time
spin-$1/2$ Weyl field.

Entirely similar observations hold for the $K=8$
fractional superstring. This string has critical
dimension $D=4$, conformal anomaly $c_0=12/5$ per
space-time dimension, and a fractional supercurrent
$G^K$ of dimension $6/5$.  For the $K=8$ closed
fractional superstring, we find one tachyon-free
modular-invariant partition function
  \begin{equation}
  {\cal Z}_{(8,8)}(q)= |A_8|^2+|B_8|^2+2|C_8|^2,
  \end{equation}
where
  \begin{eqnarray}
  A_8 = && 2(c^0_0+c^8_0)(c^2_0+c^6_0)-2(c^4_4)^2
   +8(c^0_4c^2_4)-2(c^4_0)^2,\nonumber\\
  B_8 = && 4(c^0_0+c^8_0)(c^2_4)+4(c^2_0+c^6_0)(c^0_4)
   -4(c^4_0c^4_4),\nonumber\\
  C_8 = && 4(c^2_2+c^6_2)(c^0_2+c^8_2)-4(c^4_2)^2.\nonumber
  \end{eqnarray}
The partition function blocks mix under the $S$ modular
transformation as $S(A_8)=(1/2)A_8+(1/2)B_8+C_8$, $S(B_8)=
(1/2)A_8+(1/2)B_8-C_8$ and $S(C_8)=(1/2)A_8-(1/2)B_8$.
Eq.~(\ref{sjm}) implies that $A_8\sim q^0(1+\ldots)$,
$B_8\sim q^{1/2}(1+\ldots)$, and $C_8\sim q^{3/4}(1+\ldots)$.
The massless states only contribute to the terms
$2c^0_0c^2_0-2(c^4_4)^2$ in $A_8$.  Again, the $q^0$ term in
$2c^0_0c^2_0$ can be identified with the massless vector
state of Eq.~(\ref{Vect}).  The term $2(c^4_4)^2$ must
be interpreted as space-time fermions.

The massless fermion states of the fractional superstrings
are described in a manner closely analogous to those of the
$K=2$ superstring.  They appear in the above partition functions
in terms of the form $(c^{K/2}_{K/2})^{D-2}$.  The spin field
$\sigma_{K/2}$ is the lowest dimension parafermion field in
$c^{K/2}_{K/2}$.  We write the fermion ground state as
  \begin{equation}
  |\phi\rangle=\prod^D_{\mu=1}\sigma^\mu_{K/2}|p\rangle,
  \end{equation}
in analogy to the Ramond ground-state of the superstring.
The energy operator $\epsilon^\mu$ can have integer moding
when acting on $\sigma^{\mu}_{K/2}$.  From the
parafermion theory,\cite{ZFpara} one can show that
$\epsilon^\mu_0\epsilon^\mu_0=1$ when acting on
$\sigma^{\mu}_{K/2}$.  With appropriate Klein factors
$\epsilon^\mu_0$ satisfies the Clifford algebra,
$\{\epsilon^\mu_0,\epsilon^\nu_0\}=-2g^{\mu\nu}$,
when acting on $|\phi\rangle$, where $g^{\mu\nu}$
is the Minkowski metric.  Thus the $\epsilon^\mu_0$ can be
identified with Dirac gamma matrices and $|\phi\rangle$
has the degeneracy and Lorentz properties of a spinor.
$|\phi\rangle$ is massless if we choose intercept
zero for this sector:  $L_0|\phi\rangle=p^2|\phi\rangle=0$.
The only non-trivial physical state condition on $|\phi\rangle$
is given by the $G^K$ zero mode $G^K_0=\sum_n(\alpha^\mu_{-n}+
c^\prime_n\epsilon^\mu_{-n})\epsilon_{n,\mu}$.
We find for the physical state condition $0=G^K_0|\phi\rangle
=p\cdot\epsilon_0|\phi\rangle=\not{\hbox{\kern-2.3pt $p$}}
|\phi\rangle$, giving the massless Dirac equation.  With a
Weyl and/or Majorana projection, the number of physical spinor
degrees of freedom is $8$, $4$, and $2$ in $10$, $6$, and $4$
dimensions, respectively, in agreement with the ${\cal Z}_{(K,K)}$
partition functions.

Note that the massless spectrum in the closed string partition
function ${\cal Z}_{(K,K)}$ has the correct counting of states
for a $D$-dimensional $N=2$ supergravity theory.
The existence of massless
spin-$3/2$ states suggests that if the theory is to be unitary,
it must have space-time supersymmetry.  This implies that
${\cal Z}_{(K,K)}$ vanishes, which in turn implies that
$A_K=B_K=C_K=0$.  We have checked, using known expressions
for the string functions,\cite{GQ} that $A_K$, $B_K$ and $C_K$
are each identically zero out to the first ten thousand
terms in their expansions in $q$.  It turns out that these
identities can indeed be proven.\cite{Keith}

In summary, there exist modular invariant combinations
of the string functions which remove tachyon contributions
by projections similar to the GSO projection.\cite{GSO}
The resulting partition functions have only
positive and negative integer coefficients, permitting
a space-time particle interpretation which is supported
by the explicit construction of the massless degrees of
freedom.
Finally, the partition functions obey an exact cancellation
of boson and fermion degrees of freedom at each mass
level, indicative of space-time supersymmetry.

The arguments presented in this letter also suggest
other types of new string theories:
(1) So far we have built only closed string
theories in which the left- and right-moving
world-sheet symmetries are matched.  However, one
can also build new types of heterotic strings in
which the left- and right-moving worldsheet
symmetries are different.\cite{Keith}  (2) Another
possible new string theory has $K=16$ and critical
dimension $D=3$ which could describe a theory
of space-time anyons.  Since there is only one
transverse direction, the modular invariant partition
functions ${\cal Z}_{(16,16)}$ have been classified.
All three solutions in the ADE classification\cite{CIZ}
have tachyons.  However, it is still possible that a
heterotic $K=16$ string could be tachyon-free.
(3) A final possiblity is the $K=\infty$ string with
$D=2$ where $G^\infty$ has unit dimension. It has partition
function ${\cal Z}_{(\infty,\infty)}=$constant.
We believe such a string can be constructed by gauging
appropriate WZW currents on the string world-sheet.

There are clearly many issues that must be addressed in
order to show that the fractional superstrings proposed
here are consistent theories.  Foremost among these are
the questions of unitarity, scattering
amplitudes, the ghost system and the
local world-sheet symmetry underlying the fractional
supersymmetry constraint algebra.  The techniques that
answer these questions for the $K=2$ superstring depend
largely on the fact that the superVirasoro algebra is
a local algebra on the world-sheet.  The $K=4$ fractional
supersymmetry chiral algebra is nonlocal; however there
exists a splitting of its current that satisfies an
algebra with Abelian (parafermionic) braiding relations,
allowing a description of the algebra in terms of generalized
commutators of modes.\cite{FZft}  We believe that the $K=8$
and $16$ algebras are both nonlocal and non-Abelian,\cite{AGT}
and will require new techniques to approach them.  The fractional
superstrings proposed here thus present conceptual and
calculational challenges; they also promise rich and novel
worldsheet structures.

It is a pleasure to thank K. Dienes, J. Grochocinski,
A. LeClair, and E. Lyman for discussions.  This work
was supported in part by the National Science Foundation.



\begin{thebibliography}{99}

\bibitem{GSW}See {\it e.g.} M.B.~Green, J.H.~Schwarz, and
E.~Witten, {\it Superstring Theory}, Cambridge University Press
(1987).

\bibitem{KMQ}D.~Kastor, E.~Martinec and Z.~Qiu, Phys. Lett.
{\bf 200B} (1988) 434.

\bibitem{BNY}J.~Bagger, D.~Nemeschansky and S.~Yankielowicz,
Phys.  Rev. Lett. {\bf 60} (1988) 389; F.~Ravanini, Mod. Phys.
Lett. {\bf A3} (1988) 397.

\bibitem{ZFpara}A.B.~Zamolodchikov and V.A.~Fateev, Sov.
Phys. J.E.T.P. {\bf 62} (1985) 215.

\bibitem{FZft}A.B.~Zamolodchikov and V.A.~Fateev, Theor. Math.
Phys. {\bf 71} (1987) 451.

\bibitem{Pog}R.~Poghossian, Int. J. Mod. Phys. {\bf 6A}
(1991) 2005.

\bibitem{CLT}S.~Chung, E.~Lyman and S.-H.H.~Tye, Cornell
preprint CLNS 91/1057, {\it ``Fractional Supersymmetry and
Minimal Coset Models in Conformal Field Theory.''}

\bibitem{AGT}P.C.~Argyres, J.~Grochocinski and S.-H.H.~Tye,
Cornell preprint CLNS 91/1059, {\it ``Structure Constants of
the Fractional Supersymmetry Chiral Algebras.''}

\bibitem{ALT}The tensor product algebra is not equivalent
to the single component algebra, and so leads to different
critical central charges than those found in P.C.~Argyres,
A.~LeClair and S.-H.H.~Tye, Phys. Lett. {\bf 253B} (1991) 306.

\bibitem{BrN}L.~Brink and H.B.~Nielsen, Phys. Lett.
{\bf 45B} (1973) 332.

\bibitem{ArM}A comparison of the critical central charges
shows that the fractional superstrings proposed here are
unrelated to the parastring theories of F.~Ardalan and
F.~Mansouri, Phys. Rev. Lett. {\bf 56} (1986) 2456; Phys. Lett.
{\bf B176} (1986) 99.

\bibitem{GQ}V.G.~Kac, Adv. Math. {\bf 35} (1980) 264;
V.G.~Kac and D.~Peterson, Bull. AMS {\bf 3} (1980) 1057;
Adv. Math. {\bf 53} (1984) 125;
D.~Gepner and Z.~Qiu, Nucl. Phys. {\bf B285} (1987) 423.

\bibitem{GSO}F.~Gliozzi, J.~Scherk and D.~Olive, Nucl.
Phys. {\bf B122} (1977) 253.

\bibitem{taut}Throughout this letter, the appropriate powers
of ${\rm Im} \tau$ in partition functions and of $\tau$
in their modular transformations are suppressed.

\bibitem{Keith}K.R.~Dienes and S.-H.H.~Tye,
in preparation.

\bibitem{CIZ}A.~Cappelli, C.~Itzykson and J.-B.~Zuber, Nucl.
Phys. {\bf B280} (1987) 445.

\end{thebibliography}
\end{document}